# A moderately precise dynamical age for the Homunculus of Eta Carinae based on 13 years of HST    imaging

Nathan Smith[1]*

[1]Steward Observatory, University of Arizona, 933 N. Cherry Ave., Tucson, AZ 85721, USA

19 June 2017

ABSTRACT

The *Hubble Space Telescope* (*HST*) archive contains a large collection of images of η Carinae, and this paper analyzes those most suitable for measuring its expanding Homunculus Nebula. Multiple intensity tracings through the Homunculus reveal the fractional increase in the overall size of the nebula; this avoids registration uncertainty, mitigates brightness fluctuations, and is independent of previous methods. Combining a 13-yr baseline of Wide Field Planetary Camera 2 (WFPC2) images in the F631N filter, with a 4-yr baseline of Advanced Camera for Surveys/High Resolution Channel (ACS/HRC) images in the F550M filter, yields an ejection date (assuming linear motion) of 1847.1 (±0.8 yr). This result improves the precision, but is in excellent agreement with the previous study by Morse et al. (2001) that used a shorter time baseline and a different analysis method. This more precise date is inconsistent with ejection during a periastron passage of the eccentric binary. Ejection occured well into the main plateau of the Great Eruption, and not during the brief peaks in 1843 and 1838. The age uncertainty is dominated by a real spread in ages of various knots, and by some irregular brightness fluctuations. Several knots appear to have been ejected decades before or after the mean date, implying a complicated history of mass-loss episodes outside the main bright phase of the eruption. The extended history of mass ejection may have been largely erased by the passage of a shock through clumpy ejecta, as most material was swept into a thin shell with nearly uniform apparent age.

## 1 INTRODUCTION

The bipolar Homunculus Nebula around $\eta$ Carinae provides a stunning reminder that steady winds are not the only way stars shed mass, and that the late evolution of massive stars can be punctuated by brief but violent episodes of unstable, eruptive mass loss (Smith & Owocki 2006; Smith 2014). High-resolution *Hubble Space Telescope* (*HST*) images of the Homunculus (e.g., Morse et al. 1998 and references therein) show a complex web of knots and filaments in a dusty bipolar shell with an unusual equatorial skirt. The bipolar lobes are extremely thin, hollow, and dense molecular structures ejected by the star (Smith 2002; Smith et al. 2003; Smith 2006; Nielsen et al. 2005). While observations have constrained many of the most basic properties of the Homunculus (see the review by Smith 2012 and references therein), understanding its origin and shaping remains an enduring challenge.

The $\eta$ Car system is a highly eccentric binary with a $\sim$5.5 yr orbital period, and with signatures of strong wind interactions (Damineli 1996; Damineli et al. 1997; Corcoran et al. 2001; Pittard & Corcoran 2002; Parkin et al. 2011; Madura et al. 2012). It is widely suspected that binary interaction plays some role in the Great Eruption, but the level of involvement (i.e. an agent that modifies the outflow geometry, a trigger of the instability, or a driving agent and power source) remains difficult to ascertain.

The mass ejection history inferred from the motions of nebular material has important bearing on interpreting the nature of $\eta$ Car. In a recent study, Kiminki et al. (2016) measured proper motions of the extended Outer Ejecta, showing strong evidence for multiple major mass ejections several centuries before the Great Eruption. Proper motions for features deep inside the Homunculus indicate mass ejections many decades after the Great Eruption as well (Dorland et al. 2004; Smith et al. 2004a; Artigau et al. 2011).

The present paper concentrates on the bipolar Homunculus, which is more closely related to the historical light curve of $\eta$ Car's 19th century Great Eruption (Smith & Frew 2011), and informs our interpretation of the light curves  of potentially similar extragalactic transients now being discovered (Van Dyk & Matheson 2012; Smith et al. 2011). The historical light curve of $\eta$ Car shows a complicated series of brief brightening episodes occurring in 1838 and 1843 (and perhaps earlier in 1827), and then a longer duration bright phase in 1845 until the late 1850s (Figure 1). The eruption may result from instability near the Eddington limit (Langer et al. 1994, 1999). It has traditionally been thought that super-Eddington luminosities lifted material off the star in a continuum-driven wind (Owocki et al. 2004; Owocki & Shaviv 2016), but during which one (or more) of these luminosity spikes was the Homunculus ejected? In a somewhat different viewpoint, there are several clues pointing to an explosive nature of the eruption (Smith 2006, 2008; Smith et al. 2003), in which case a significant fraction of the eruption luminosity may come from shock interaction with circumstellar material (Smith 2013). Shock structures from older eruptions are also evident in the Outer Ejecta (Kiminki et al. 2016; Mehner et al. 2016).

* E-mail: nathans@as.edu

The plot thickens, because some of the peaks in the light curve coincide with times of periastron in the eccentric binary system (Damineli 1996; Smith & Frew 2011). Smith (2011) showed that for the observed color and brightness of $\eta$ Car in the time period leading up to the Great Eruption, the photospheric radius was larger than the separation of the stars at periastron. This means that there must have been a direct violent collision of some sort between the two stars of the currently observed binary at times of periastron, and that the currently observed binary orbit was similar before the eruption. (Note that the coincidence of periastron with the pre-1845 luminosity spikes requires that the orbital period was ~5% shorter than presently observed, which is appropriate because the ejection of the Homunculus removed ~10% of the system's total mass; Smith & Frew 2011.) One would obviously like to know if the mass contained in the Homunculus was ejected in one of these brief periastron luminosity spikes in 1843, 1838, or earlier. Alternatively, it may have been ejected at some later time during the main plateau of the Great Eruption in 1845 till the late 1850s. The answer to this question helps determine if the brief periastron encounters were the main event, or a prelude to something more extreme.

A new window for understanding $\eta$ Car's eruption has opened with the discovery of its light echoes (Rest et al. 2012). Study of these echoes is underway (Rest et al. 2012; Prieto et al. 2014; Smith et al. 2017a,b), and will eventually provide a detailed spectroscopic time sequence of the 19th century eruption. This makes $\eta$ Carinae analogous to some historical Galactic supernova (SN) remnants, where the discovery of SN light echoes (Rest et al. 2005a,b, 2006, 2008, 2011a,b; Krause et al. 2008; Sinnott et al. 2013; Finn et al. 2016) provides a way to connect the historical event to the presently observed expanding remnant. This link has been extremely valuable for understanding explosion physics, because SN remnants allow us to dissect the shrapnel of the explosion (see, e.g., Milisavljevic & Fesen 2015) to a level of detail that is not possible with transients in distant galaxies. They provide quantitative constraints on the mass, kinematics, total kinetic energy, shock structure, geometry, composition, chemistry, and other key physical properties of the explosion. Similarly, the Homunculus allows detailed estimates of these quantities for $\eta$ Car. Light echoes provide a meaningful way to link the detailed physical properties of historical SN remnants to observations of far more plentiful extragalactic SNe, where a detailed record of the spectral evolution can be recorded with modern instruments. By the same token, $\eta$ Car provides potential links to modern extragalactic SN impostors.

In the case of $\eta$ Carinae, however, connecting the spectral evolution that is encoded in these light echoes to both the historical light curve of the Great Eruption and the properties of the Homunculus is more challenging than for historical SNe and their echoes. Historical SNe correspond to a single brief brightening event that is well-timed and undoubtedly coincident with the time of explosion (after a brief rise time that is fairly well known for normal SN types). For $\eta$ Car, the problem is more complex due to the multiple peaks in an eruption that spanned decades, an incomplete historical record of these multiple peaks (especially at early times), and an uncertain time delay for each individual echo. The difficulty in interpreting these echo spectra can be mitigated, however, if we can infer times in the historical light curve when the most important mass ejection occured.

The most straightforward way to estimate the time of the dominant mass ejection in the Great Eruption is to directly measure the expansion age of the Homunculus Nebula via its proper motion. Although previous proper motion studies agree on the basic fact that the Homunculus was ejected sometime during the 15-20 yr Great Eruption, they disagree on a precise date. Early investigations using ground-based images were roughly consistent with an ejection during the Great Eruption, although with low precision (Thackeray 1949; Gaviola 1950; Ringuelet 1958; Gehrz & Ney 1972). The advent of *HST* imaging offered a new avenue to investigate this question. Using early epochs of pre-costar Wide Field / Planetary Camera (WF/PC) images and WFPC2 images, and taken in different filters that contained different emission lines, spread over a 2 yr baseline, Currie et al. (1996) measured an ejection date of 1841.2 ($\pm 0.8$ yr)[1]. However, Morse et al. (2001) measured a later date of 1847.4 ($\pm 5$ yr) using only WFPC2 images with a 4-5 yr baseline in consistent filters. Independently, Smith & Gehrz (1998) found 1843.8 ($\pm 7.3$ yr) using ground-based images obtained over 50 yrs (including an *HST* image smoothed to ground-based resolution, but with ground-based images that had different wavelength sensitivity). These various studies (see Figure 1) do not fully agree, and do not provide the precision needed to confidently determine if the Homunculus was ejected in the 1838 event, the 1843 event, or sometime during the longer brightening episode beginning around 1845 till the late 1850s.

A more precise constraint on the dynamical age is of interest, however, because the differences in these are significant in light of more recent studies – especially for the interpretation of light echoes, as noted above. For example, spectra of light echoes from early peaks showed evidence of rapid molecule formation (Prieto et al. 2014), and the Ho-

---

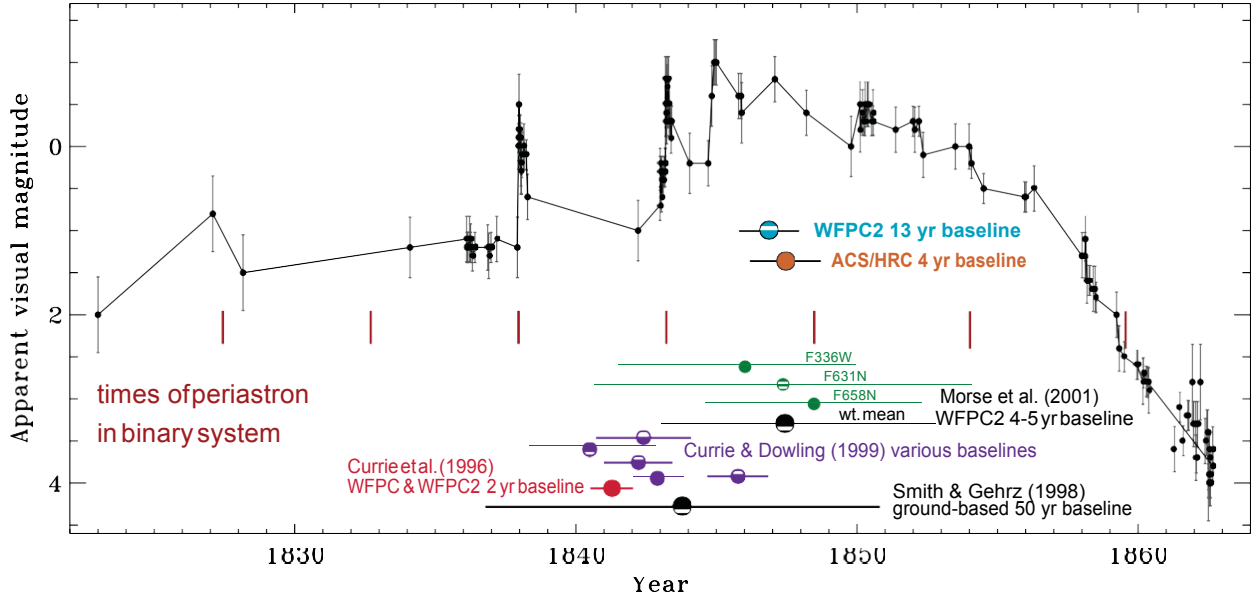

Figure 1. The revised historical light curve of the 19th century Great Eruption of η Car from Smith & Frew (2011). The red hash marks show expected times of periastron in the eccentric binary system, extrapolating from the orbital cycle observed in modern times (see Smith & Frew 2011). On the bottom we show previous estimates for the ejection dates of the Homunculus. Proper motions measured with WF/PC and WFPC2 images over a ~2 yr baseline (Currie et al. 1996; Currie & Dowling 1999) are shown in red and purple, respectively, with WFPC2 over a 4-5 yr baseline (Morse et al. 2001) are shown in black (mean) and green (individual filters), and with ground-based images taken over a 50 yr baseline (Smith & Gehrz 1998) in black. The results of the present study using WFPC2 and ACS are also shown.

munculus includes dense molecular gas (Smith 2002, 2006; Smith et al. 2006; Smith & Ferland 2007; Loinard et al. 2012, 2016). However, the most recent proper motions measured by Morse et al. (2001) point to a later ejection date. Therefore, this paper revisits the proper motion of the Homunculus using available archival *HST* images. The available WFPC2 images include sufficiently long exposures in the same F631N filter that span 13 years. This increased time baseline should significantly improve upon the precision in ejection date as compared to Morse et al. (2001), which had only a 4-5 yr baseline. As it turns out, the limitations in the precision of the ejection date are driven by a real spread in ages for individual condensations, plus confusion between motion and variable illumination of the ejecta. A complementary recent paper presented proper motion measurements for the older, Outer Ejecta around η Car (Kiminki et al. 2016), finding evidence for multiple previous eruptions spanning 600 yr, including some material ejected decades preceding the 1840s eruption.

## 2 ARCHIVAL IMAGING AND ANALYSIS

There is a wide assortment of high-quality imaging of η Car and its surroundings taken with *HST* over the past 2 decades, using several different instruments and filters. Many of these images that populate the archive are poorly suited for measuring proper motions of the Homunculus because of exposure times that are too long (heavily saturated) or too short (poor signal-to-noise ratio in the polar lobes), or because they are heavily contaminated by emission lines that are time-variable or spatially variable in some filters, such as F658N. The two instrument configurations chosen for anal-ysis here are the F631N filter of WFPC2, and the F550M filter of the Advanced Camera For Surveys High-Resolution Channel (ACS/HRC). Overviews of the data quality and characteristics of η Car images were discussed by Morse et al. (1998) for WFPC2 and by Smith et al. (2004a) for ACS/HRC. The spatial dependence of the time variability in images of the Homunculus was discussed previously by Smith et al. (2000) for WFPC2 and Smith et al. (2004b) for ACS/HRC, including descriptions of how extended emission lines cause substantial changes during the 5.5 yr cycle of η Car. Although the image quality with the F336W filter on WFPC2 and the F250W filter on ACS/HRC is somewhat sharper than F631N or F550M because of the better diffraction limit at shorter wavelengths, these filters are heavily contaminated by a forest of emission lines that contribute to the time-variable "Purple Haze" (Smith et al. 2004b). Since the Purple Haze is known to have strongly time-variable line emission, these filters are not used for proper motions here.

A summary of the imaging data used for analysis is provided in Table 1. These images were obtained and the initial reduction steps were done in basically the same way as for earlier imaging. Data reduction steps for WFPC2 imaging of η Car were discussed by Morse et al. (1998) and for ACS/HRC imaging by Smith et al. (2004a). There are a few basic differences in the present study that are relevant to measuring proper motions. First, all the data were downloaded from the archive with the most recent distortion corrections applied. Second, these previous studies used a range of exposure times to cover the extreme dynamic range between the bright central star and the fainter polar lobes; this required one to use short exposures of the unsaturated central star to patch in saturated regions of longer exposures.

Table 1. Archival *HST* Images of $\eta$ Car

| Dataset | Instrument | Filter | Date (start) | Exp.(s) |
|---------|-----------|--------|--------------|---------|
| U2O80105T | WFPC2/PC1 | F631N | 1995 09 21 | 30.0 |
| U3GN0109M/AM | WFPC2/PC1 | F631N | 1997 06 08 | 40.0 (×2) |
| U5JE010FR/GR | WFPC2/PC1 | F631N | 1999 06 12 | 40.0 (×2) |
| U6CK010FR/GR | WFPC2/PC1 | F631N | 2001 06 04 | 40.0 (×2) |
| U8GM1W09M/FR | WFPC2/PC1 | F631N | 2003 02 12 | 20.0 (×2) |
| U8MA4W09M/FR | WFPC2/PC1 | F631N | 2003 07 19 | 20.0 (×2) |
| U8PB0103M/JM | WFPC2/PC1 | F631N | 2003 08 07 | 20.0 (×2) |
| UB7C0103M/JM | WFPC2/PC1 | F631N | 2008 09 05 | 20.0 (×2) |
| J8GM1A0A0 | ACS/HRC | F550M | 2002 10 14 | 9.6 |
| J8GM2A0F0 | ACS/HRC | F550M | 2003 02 12 | 40.0 |
| J8MA3A0F0 | ACS/HRC | F550M | 2003 06 13 | 40.0 |
| J8MA4A0F0 | ACS/HRC | F550M | 2003 07 20 | 40.0 |
| J8MA5A0F0 | ACS/HRC | F550M | 2003 09 13 | 40.0 |
| J8MA6A0F0 | ACS/HRC | F550M | 2003 11 14 | 40.0 |
| J8MA7A0F0 | ACS/HRC | F550M | 2004 12 06 | 40.0 |
| J8MA8A0K0 | ACS/HRC | F550M | 2005 07 14 | 40.0 |
| J8MA9A0K0 | ACS/HRC | F550M | 2005 11 06 | 40.0 |
| J9P602080 | ACS/HRC | F550M | 2006 08 04 | 42.0 |

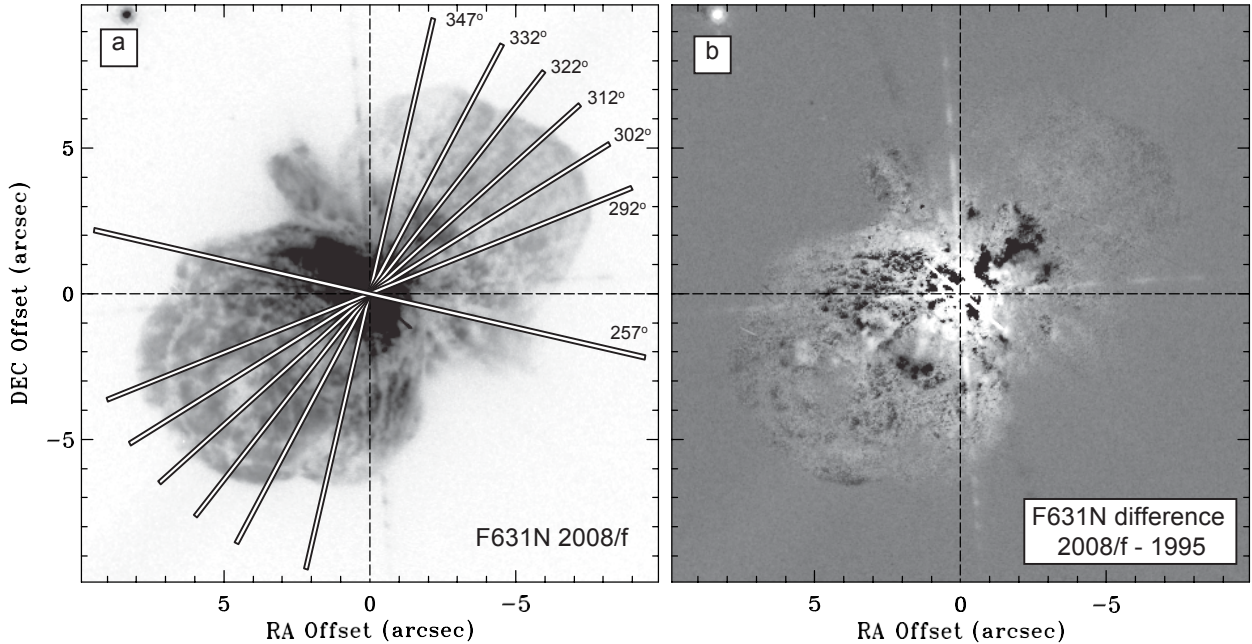

Figure 2. (a) *HST*/WFPC2 F631N image of the Homunculus in 2008 (shrunk to match the 1995 image size scale), showing the position angles of the seven intensity tracings described in this paper. (b) A difference image made by subtracting the 1995 F631N image from the 2008 image that was reduced in size by a factor $f = 1.0872$. The residuals are due mainly to non-uniform changes in brightness, rather than positional shifts. Black features were relatively brighter in 1995, and white features were brighter in 2008. Note that in the 2008 image, the Homunculus was brighter overall by 34%, and so the flux in the 1995 images was scaled up to match this before subtraction in order to emphasize non-uniform changes in flux. See the online edition for an animated gif that blinks the two frames used to make the subtraction in panel (b). (Referee: these are appended at the end of the PDF file, with the 1995 image first and the 2008 image second.)

Usually, this is done in several steps to patch ever increasing numbers of pixels affected by saturation and CCD blooming as exposure times are increased. Each step might introduce minor registration errors between exposures. Instead, we simply downloaded individual frames with appropriate signal to noise in the polar lobes of the Homunculus, and ignored the bleeding of the saturated central star. The third major difference is that we did not perform Lucy-Richardson deconvolution on the images, as had been done previously (Morse et al. 1998, 2001; Smith et al. 2000, 2004a,b). Finally, the last difference is that in previous studies of the proper motion and variability by Morse et al. and Smith et al., images in different filters and from multiple epochs were registered using a handful of faint surrounding field stars, requiring a geometric transformation (using geomap in IRAF) to the image frames, which introduces a small amount of distortion to the images. This geometric transformation was not performed in the present study, because we do not register the star's position in the images; we only measure the increase in size across the Homunculus ignoring the position of the star.

The primary goal of this paper is to provide a quick and reliable solution to the discrepancy in previous estimates of the age of the Homunculus from proper motions, which as noted in the Introduction, are given as 1841, 1843, and 1847 (Currie et al. 1996; Smith & Gehrz 1998; Morse et al. 2001). A resolution of these differences is needed for interpreting spectroscopy of $\eta$ Car's light echoes, which correspond to various individual peaks in the historical light curve (there are many additional light echoes still being studied, beyond those already discussed by Rest et al. 2012; Prieto et al. 2014). It has been more than 15 years since the last of these previous proper motion studies of $\eta$ Car's Homunculus, and the *HST*/WFPC2 imaging time baseline has increased, while we also have the addition of multiple epochs of ACS/HRC data with smaller pixels.

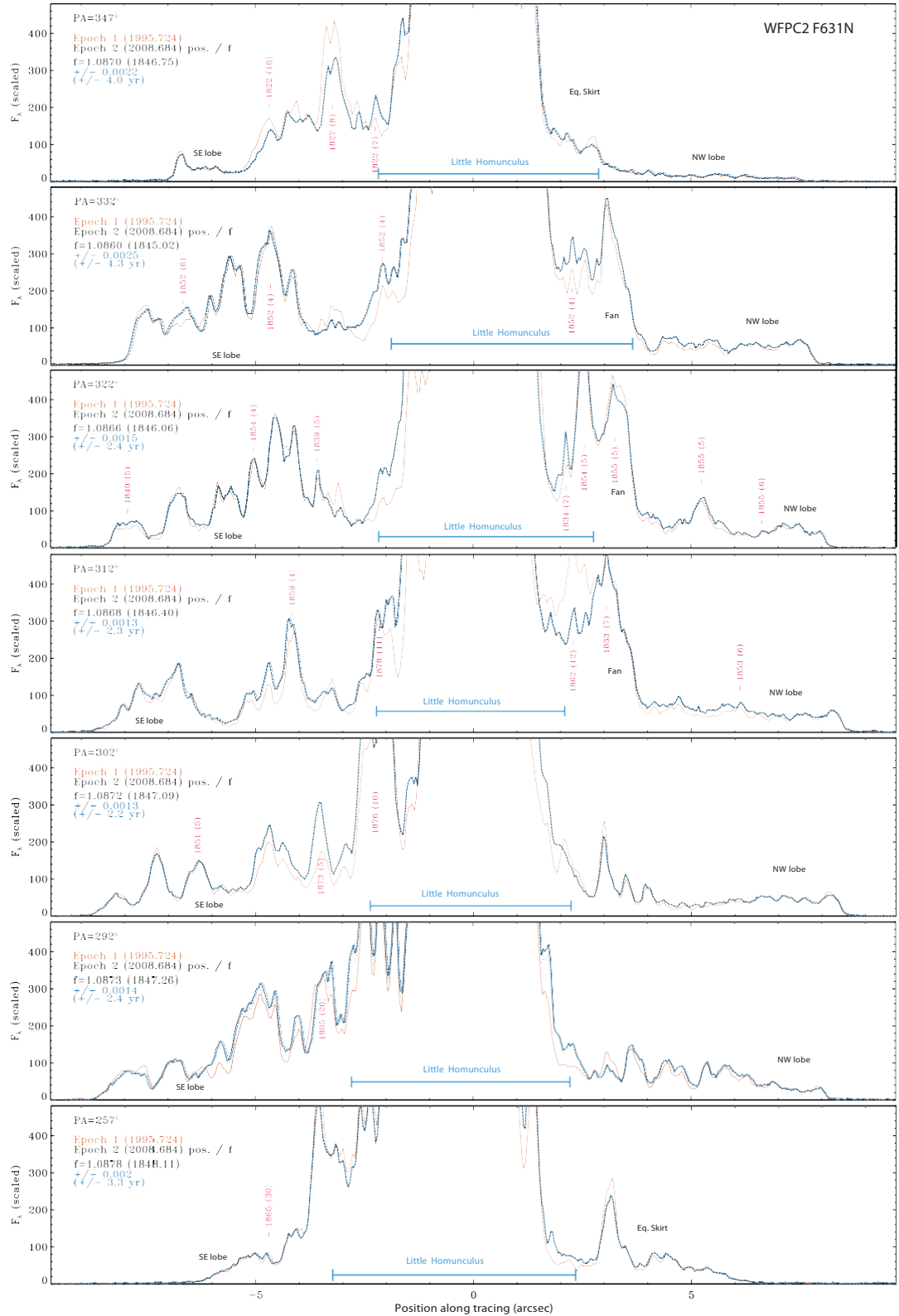

**Figure 3.** Linear intensity tracings across the Homunculus at the seven position angles shown in Figure 2a. Epoch 1 is in orange, and Epoch 2 (black) has been shrunk by a factor *f* to match the size scale of Epoch 1. The blue intensity tracings show Epoch 2 shifted by ±1 σ. Epoch 1 has been scaled up in flux by 34% to match the average brightness in the polar lobes. The factor *f* and the corresponding dynamical age are indicated in each panel. Features that have widely discrepant motion have their implied ejection ages shown in magenta with an approximate error in years in parentheses. For each position angle, the light blue bar at the bottom shows the extent of the Little Homunculus (LH) as determined by Smith (2005).

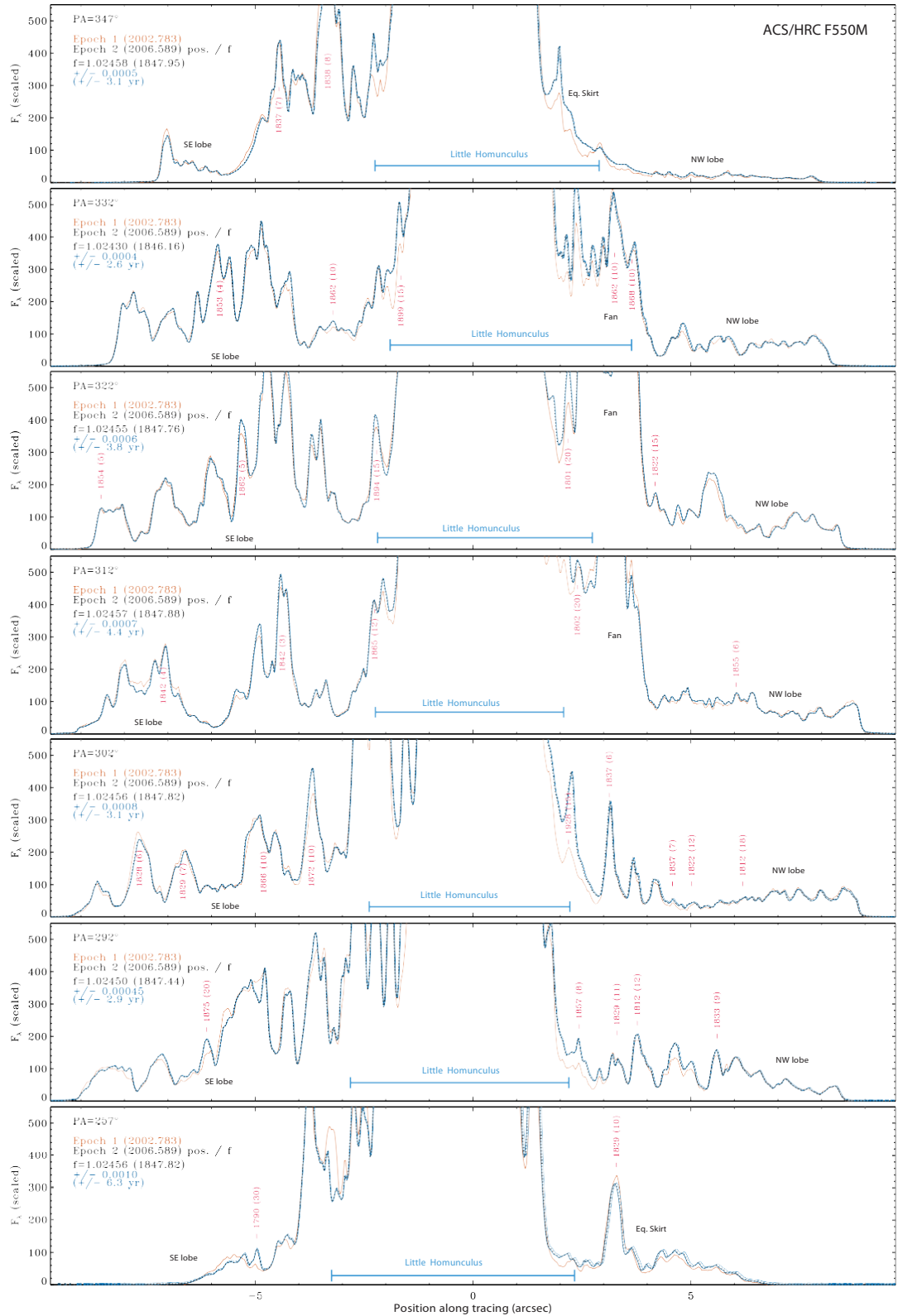

**Figure 4.** Same as Figure 3, but for the ACS/HRC images in the F550M filter.

Since our goal is to provide an independent and quick estimate of the average age of the Homunculus, we take a somewhat different approach from previous studies. Currie et al. (1996) and Morse et al. (2001) both used a method similar that used to measure proper motions of protostellar Herbig-Haro jets (Hartigan et al. 2001; Bally et al. 2006; Reiter & Smith 2014; Reiter et al. 2015), wherein small boxes are selected to encompass individual knots or groups of knots and then the cross correlation is measured for each box. For $\eta$ Car, this technique has a few potential systematic effects: it may be hindered by the limiting precision of registering images, uncertainty in the position of the central star, the distortion corrections needed to align the images, and strong brightness gradients or temporal variability in brightness across the Homunculus. The measurement technique adopted here is not necessarily superior to the cross correlation of indiviudual knots used previously, but it is chosen to avoid these possible systematic effects and thereby provide an independent check. Indeed, over the same time baseline (using the same images) as Morse et al., the analysis used here has lower precision because the deconvolved images used by Morse et al. are sharper (although the motion we measure on that shorter baseline is consistent within the uncertainty). The real advantge of this study in precision comes from the longer time baseline, and the inclusion of the ACS images that provide a fully independent dataset.

This study aims to measure the fractional increase in the total extent of the nebula. To some extent, the answer one gets may depend on which features one choses to measure. The outer portions and edges of the polar lobes of the Homunculus are the fastest features, and contain most of the power to constrain the age. The outer portions of the polar lobes are also the least affected by strong non-uniform brightness variations due to the central star's photometric variability (Smith et al. 2000), or by motion and variability of the younger Little Homunculus (LH) inside the core of the Homunculus (Ishibashi et al. 2003; Smith 2005). We also focus on the longest time baselines of images with consisent filters, although we examined intermediate epochs to check for linear motion. The average age derived from all baselines are consistent with linear motion (i.e., constant speed) within the uncertainty, as found already by (Morse et al. 2001), although we did not delve into a careful analysis of the linearity of every individual condensation, because this is far beyond the main goal of constraining the average age. Such non-linear motion of individual knots may exist, and may be worthwhile to investigate in the future.

Our method to determine the age is to evaluate the over-all scale factor $f$ by which the Homunculus has increased in size ($S$) between two epochs (.i.e. $S_2 = f \times S_1$). One way we do this is to produce linear intensity tracings across the Homunculus polar lobes, passing through the star at the seven different position angles shown in Figure 2a. Overplotting two epochs of tracings at the same position angle, we then calculate the optimum scale factor $f$ by which the Epoch 2 image must be shrunk in order to align with structures in the Epoch 1 image (applying an arbitrary lateral shift to align the images). For images at Epoch 1 ($t_1$) and Epoch 2 ($t_2$), the ejection date ($t_0$) is then given by

$$t_0 = \frac{f t_1 - t_2}{f - 1} \qquad (1)$$

where values of $t$ are decimal years. It became immediately apparent, however, that there was no straightforward choice for a best shift, because there were always large residuals in different parts of the nebula. Upon close inspection of the individual tracings, many features were aligned well based on their centroids, but were leaving large residuals after subtraction because they had actually changed brightness substantially. While there is an overall brightening of the Homunculus from 1995 to 2008 of about 34%, this has been normalized. The brightness changes that persist after this normalization are rather complicated. Some condensations brightened by large factors while other condensations faded during the same time interval. This complicates the situation. The dramatic changes in brightness were most severe in the inner portions of the Homunculus, overlapping with the locations of the Purple Haze (Smith et al. 2004a) and the LH (Ishibashi et al. 2003; Smith 2005), as well as the central star, but there were some brightness variations in the outer parts as well. During the 1990s, the central star is known to have brightened dramatically (Davidson et al. 1999), and the spatially resolved pattern of brightening in the inner Homuncuus was shown by Smith et al. (2000). Figure 2b shows a difference image made by subtracting the 1995 F631N image from a 2008 F631N image that was scaled to the same size (see below). The irregular pattern of brightening and fading in the inner Homunculus is quite evident, while most of the outer regions of the polar lobes show a cleaner subtraction.

A choice was therefore made that the optimal scale factor $f$ is the one that minimizes the subtraction residual in the outer portions of the polar lobes, especially near the edges, ignoring the strong fluctuations in the inner nebula. Figures 3 and 4 show the results of shifting Epoch 2 (blue) to match Epoch 1 (orange) for both WFPC2/F631N and ACS/HRC F550M images, respectively, at each of the seven different position angles. Values of $t_1$, $t_2$, the optimal scale factor $f$, and the corresponding ejection dates $t_0$ are noted in each panel, and listed in Table 2. When tracings are aligned less optimally, the subtraction residuals increase mainly at locations where there is a strong gradient in the flux (i.e. at boundaries of bright condensations), resulting in a pattern of both positive and negative residuals on opposing sides of a condensation aligned in the direction of motion. Uncertainty was assigned based on shifts where these subtraction residuals at edges of features increased by a factor of 2 above the noise in the subtraction. These ±$1\sigma$ shifted tracings are shown with dashed blue curves in Figures 3 and 4. This is probably a generous error bar, since such shifts were worse by eye than the optimal shift. Combining the values for all seven position angles, we find weighted mean ejection dates of 1846.8 (±1.1 yr) for the WFPC2 F631N images across a 13 yr baseline, and 1847.4 (±1.3 yr) for the F550M ACS/HRC images with a 4 yr baseline. The weighted mean of the two filters together is 1847.1 (±0.8 yr).

We performed a similar exercise on the 2-D images themselves, shrinking the Epoch 2 image by a factor $f$ and aligning it with the Epoch 1 image. The best way to confirm the optimal alignment was watching the aligned images blink on a computer screen, where the difference between motion and flux variations became obvious. This can also be seen in the difference image mentioned earlier, seen in Figure 2b (the electronic edition includes a blinking comparison in an animated gif file). The alignment and subtraction of the 2-D images confirmed the results of the 1-D intensity tracings, but were less precise due to the added registration uncertainty; once again, we made the decision to ignore the inner regions of the Homunculus and aimed to minimize the subtraction residuals in the outer polar lobes.

Although this analysis yields a reliable average age of the Homunculus, it has also exposed some anomalous motion. Several individual features appear to have significant motion that disagrees with homologous expansion (i.e. they move in the 1-D tracings that have been aligned). Some features had motion indicating that they were decades older or younger than the average date. These dates are labeled (magenta) in Figures 3 and 4. Features with ejection dates that are decades after the Great Eruption are located mainly at positions that overlap with the edges of the LH. Thus, in addition to seeing dramatic brightening of the LH, we may also have detected the proper motions for a few condensations associated with the LH itself. In contrast, some other features seem to be decades older than the rest of the Homunculus, having motion that points back to the 1820s. Currie & Dowling (1999) also noted a wide range of ejection dates within a few decades of the average age of the Homunculus from their proper motions, and Kiminki et al. (2016) noted a number of measured components of the Outer Ejecta that point to ejection in the early 1800s. Mehner et al. (2016) also associate some of the Outer Ejecta with the Great Eruption. From radial velocities in spectra, some equatorial features in the Homunculus seem decades older than the average age (Hartman et al. 2004; Zethson et al. 1999). From our analysis method, though, it is difficult to confidently tell the difference for an individual knot between real motion that is faster or slower than the average of the Homunculus, as opposed to two neighboring and marginally resolved knots that brightened unequally. Such uneven flux variation could artifically drag a centroid to appear like anomalous motion; this is a concern given the dramatic changes in brightness (Fig 2b). Perhaps repeating a more detailed analysis of individual condensations using deconvolved images or a longer time baseline of ACS/HRC images could help constrain the actual range of ejection dates, but that is beyond the scope of this investigation.

Table 2. Ejection dates and uncertainty for intensity tracings at various position angles (see Figs. 3 and 4)

| Trace P.A. | F631N | $\sigma$ | F550M | $\sigma$ |
|---|---|---|---|---|
| 257″ | 1848.1 | 3.3 | 1847.8 | 6.3 |
| 292″ | 1847.3 | 2.4 | 1847.4 | 2.9 |
| 302″ | 1847.1 | 2.2 | 1847.8 | 3.1 |
| 312″ | 1846.4 | 2.3 | 1847.9 | 4.4 |
| 322″ | 1846.1 | 2.4 | 1847.8 | 3.8 |
| 332″ | 1845.0 | 4.3 | 1846.2 | 2.6 |
| 347″ | 1846.8 | 4.0 | 1848.0 | 3.1 |
| Wt. Mean | 1846.8 | 1.1[a] | 1847.4 | 1.3[a] |

[a] Uncertainty on each weighted mean is given by $(\Sigma w_i)^{-1/2}$ and the individual weights are $w_i = \sigma_i^{-2}$.

## 3    RESULTS AND DISCUSSION

### 3.1    Comparison with previous measurements

Combining our WFPC2 and ACS/HRC proper motion measurements (Table 2) gives an average ejection date for the Homunculus in the first months of 1847. The weighted mean of the two filters is 1847.1 (±0.8 yr). This is in excellent agreement with Morse et al. (2001), who found 1847.4 ±5 yr (see Figure 1). The longer time baseline helps to significantly improve the precision compared to that study. As noted above, the precision appears to be limited mostly by the strong brightness fluctuations and perhaps a real spread of ages among the dense knots in the Homunculus. Since we used a different measurement technique and achieved essentially the same result, this suggests that the deconvolution and image alignment methods used by Morse et al. (2001) did not produce significant systematic offsets in the derived age.

On the other hand, our results disagree with most of the ages quoted by Currie et al. (1996) and Currie & Dowling (1999), which land several years earlier in the early 1840s (Figure 1), even though they use the same measurement technique as Morse et al. (2001). This disagreement may indicate that measuring proper motions across different filters (which transmit different emission lines) and using older pre-COSTAR WF/PC data can cause substantial systematic offsets in the measured motion.

Our derived age is consistent with the average age of the polar lobes found by Smith & Gehrz (1998), but that study used ground-based data and had a rather large error bar. Smith & Gehrz (1998) used a smoothed HST image, plus older ground-based images taken on blue-sensitive photographic plates (the original plates taken by E. Gaviola; Gaviola 1950) and color slide film (Gehrz & Ney 1972). These have different wavelength responses. This probably doesn't matter too much for the polar lobes, which are seen primarily in starlight reflected by dust, but the strong variations of the Purple Haze and LH could cause pronounced systematic effects in the inner parts of the nebula from these data. Proper motions and kinematics show that the LH and some other ejecta in the inner core of the Homunculus are indeed younger than the Great Eruption (Dorland et al. 2004; Smith et al. 2004a; Smith 2005; Artigau et al. 2011). Note that in Figure 2b, the so-called "Fan" to the north-west of the star shows some of the strongest variations in bright-ness, and color images and spectra show that it is contaminated by variable blue/violet emission lines with unusual kinematics (Hartman et al. 2004; Morse et al. 1998; Smith 2005; Smith et al. 2004b; Zethson et al. 1999). With such stark photometric variability extrapolated decades into the past, this may explain why Smith & Gehrz (1998) derived a later ejection date for this feature, which is not confirmed with proper motions in a red continuum filter (Morse et al. 2001).

### 3.2 Brightening and Fading of the Homunculus

The nonuniform pattern of brightening and fading in the inner parts of the Homunculus is interesting. The result in Figure 2b is not merely a global brightening or fading of the nebula in response to the star getting brighter (as noted earlier, there *is* such a global brightening, but this has been normalized before making the difference image), nor is it a simple matter of the star appearing brighter because the Homunculus walls are expanding and thinning (although see van Genderen & Sterken 2005). In both of those cases we would expect the dusty knots to brighten or fade together. Instead, there are irregularly distributed patches that became brighter, and other adjacent areas that faded significantly during the same time interval. Watching a blinking movie of the images (available in the online edition) creates the impression of broad shadows moving across the inner polar lobes. An interesting possibility is that the creation, destruction, or rearrangement of clumpy dust structures close to the star could cast shadows that change with time and are projected onto the screen of the Homunculus walls – somewhat like shadows cast by insects trapped inside a lantern. This has been envisioned to explain some of η Car's unresolved photometric variability in the past (van Genderen et al. 1999). Smith (2010) proposed that the near-infrared variability of η Car, which showed strong peaks around periaston events during the 1980s and 1990s (Whitelock et al. 1983, 1994, 2004), was caused by the formation of hot dust in the colliding winds of the binary system. Enhanced dust formation at periaston would be similar to the case of some colliding-wind Wolf-Rayet binaries that form dust episodically this way (Williams et al. 1990, 2001, 2009).

The strength of these infrared peaks has been diminishing with time as the central star has brightened in the late 1990s (Whitelock et al. 1983, 1994, 2004), so perhaps changes in the colliding wind properties lead to dust formation that has become less efficient (Smith 2010). This would provide less scattering of off dust in some inner regions (making them fainter) and would permit more starlight to reach larger distances, causing them to brighten. Alternatively, there may be a shift in the opening angle of the colliding-wind shock cone, in which the dust forms. The edge of the shock cone is near our line of sight to the star (Groh et al. 2010; Madura et al. 2012; Teodoro et al. 2012; Tsebrenko et al. 2013; Weigelt et al. 2016), so a widening of this shock cone, for example, could cause rather drastic brightening of the star and of many other clumps that scatter light toward us, while casting new shadows on other parts of the Homunculus. This sort of scenario was proposed already (Smith et al. 2000; Smith 2010), and was discussed further by Mehner et al. (2014) and Gull et al. (2016). Figure 2b provides additional evidence of these moving shadows (again, see the animation in the onine edition). Perhaps a more detailed mapping of the brightness changes with time is in order, but this is beyond the scope of the present paper. The main point to note here is that these moving illumination patterns may corrupt proper motion measurements if they are misinterpreted.

### 3.3 Implications and Interpretation

An ejection date for the Homunculus in early 1847 has important implications for understanding the historical light curve and its relation to the mass loss of the eruption. Knowing when most of the mass left the star is especially important for interpreting spectroscopy of light echoes from the eruption (Rest et al. 2012; Prieto et al. 2014).

Taken at face value, proper motions imply that the primary mass loss was associated with main plateau phase of Great Eruption in 1845-1858, weighted toward the earlier part. Of course, the eruption persisted for more than a decade, so it is possible that mass loss had a much longer duration than the ±1 yr uncertainty in our average ejection date. Yet, the Homunculus is a remarkably homologous outflow, with extremely thin walls and an apparently well-defined single age (Smith 2006; Morse et al. 2001). This could happen, despite actual prolonged mass loss over a decade, if the outflow speed increased dramatically as the eruption progressed. If an early phase with slow wind was followed by an explosive event, fast material could overtake slow material, sweeping it into a thin shell, and eventually coasting at a speed that looks like a single age (Smith 2013). This might also occur if the wind speed increased in a super-Eddington episode, causing internal shocks in the wind, as discussed by Quataert et al. (2016). Formation of a thin molecular shell may require a strong shock front that allows radiative cooling of the post-shock gas. Such a collision would mostly erase the details of the prolonged mass loss history, making such a scenario difficult to test with proper motions alone. Perhaps some of the clumps with anomolous motion detected here are especially dense clumps that were resilient enough to retain some of their initial kinematic information about prolonged early mass loss decades before the Great Eruption. Proper motions of some of the Outer Ejecta also show evidence for substantial mass loss in the early decades of the 19th century (Kiminki et al. 2016; Morse et al. 2001). Mehner et al. (2016) also associate some of the Outer Ejecta with the epoch of the Great Eruption.

While there is still room for a variety of mass loss scenarios, some specific hypotheses for the eruption can be constrained or ruled out by the derived ejection date. The later apparent ejection date in 1847 strongly rules out the idea that the Homunculus resulted primarily from a single ejection in 1843 or in any of the pre-1843 luminosity peaks, and it implies that these early events were a sideshow or a prelude, rather than the main event that supplied the kinetic energy and mass budget of the eruption.

The early peaks in the light curve coincide with times of periaston in the present-day eccentric binary system, extrapolated into the past (Smith & Frew 2011). Several authors have mused that something violent may have occurred at such close periaston passage in a pair of massive stars with an orbital eccentricity of $e \simeq 0.9$. Smith (2011) pointed out that the required radius of the emitting photosphere in the 1840s (between the peaks) was significantly larger than the periastron separation, meaning that the two stars must have collided multiple times. Soker and collaborators (Soker 2001, 2004; Kashi & Soker 2009) have discussed the possibility that significant mass transfer occurs at periastron, with accretion onto the secondary powering the eruption luminosity and blowing jets that shape the nebula. The results of proper motions indicate that the earlier periastron events in 1838 or 1843 were not important in driving the Homunculus shaping, although they don't rule out the possibility that similar interaction occurs in the later periastron events during the main plateau.

If those early events in 1843 and 1838 (or earlier) played an important role in the mass loss, it may have been to prepare a slowly expanding, dense, and probably asymmetric CSM into which the much faster ejecta would later collide (Smith 2013). This is consistent with light echoes studied so far (Rest et al. 2012; Prieto et al. 2014), which correspond to the early 1838/1843 luminosity spikes and show relatively slow outflow speeds of only ~200 km s$^{-1}$ in spectra, slower than the final coasting speed of the Homunculus.

If fast material crashed into slow material to form the Homunculus, as hypothesized above, then that fast material could have been ejected even later than 1847.1, because it may have decelerated upon crashing into the slow CSM. Future spectroscopy of light echoes may be able to test this hypothesis by tracing the line profile evolution throughout the eruption. Fast material would decelerate when it was swept up into a thin shell, and the resulting ballistic motion would indicate an apparent ejection time that was somewhat earlier than the actual impulse of kinetic energy. In that type of scenario with CSM interaction, it may be possible to have a sudden ejection of fast material at the periaston of 1848 or later (see Figure 1). This is difficult to confirm from available data, however. Moreover, it begs the question of why that periastron passage led to a violent explosion, while previous ones did not.

Finally, the age of the Homunculus helps to constrain some ideas about the Great Eruption resulting from a binary merger event. A binary merger provides an attractive potential explanation for the source of energy, mass ejection, and axisymmetric geometry of the Great Eruption (Gallagher 1989; Podsiadlowski 2010; Smith 2011; Portegies Zwart & van den Heuvel 2016; Smith et al. 2016). While the general idea of a merger to power the eruption may be attractive, models are difficult to reconcile with available data. A specific binary merger model in a triple system was recently proposed for $\eta$ Car by Portegies Zwart & van den Heuvel (2016). In their model, the Homunculus is the product of extreme mass loss energized by tidal interaction that occurred for many decades preceding a final merger event that produced the 1838 luminosity peak. This specific model is clearly ruled out by the measured age of the Homunculus, which is ejected much later than required in this model. The later age is, however, not necessarily a death blow to all merger models for the Great Eruption; a coalescence of two massive stars may be complicated, and our understanding of the detailed hydrodynamics in such a wild event is admittedly still quite poor. An important constraint, though, is that any merger model must account not only for the late 1847 apparent age of the Homunculus, but also the remaining secondary star in a wide eccentric orbit, and (perhaps harder to explain) multiple older eruptions in the past 700 yr. Kiminki et al. (2016) measured proper motions in $\eta$ Car's Outer Ejecta, which revealed conclusive evidence for a previous major outburst about 600 yr before the 1840s, and another about 300 yr before.

# 4  SUMMARY

Using available archival *HST* images of $\eta$ Car that substantially extend the time baseline compared to previous studies, we revisit the proper motion expansion of the Homunculus Nebula. Combining 13 years of F631N images from WFPC2 with 4 years of F550M images with ACS/HRC yields an average ejection date for the Homunculus of 1847.1 ±0.8 yr. This improves the precision significantly, but agrees well with the date of origin found by Morse et al. (2001). It is inconsistent with claimed ejection dates in the early 1840s. Using a different method of measuring the motion also provides an independent check that does not have the same potential sources of systematic error.

This more precise value of the age for the origin of the Homunculus helps to narrow the parameter space for models of the 19th century Great Eruption. It rules out any scenarios where most of the mass loss occured in the brief periastron luminosity spikes in 1843 or earlier. We caution, however, that the precision of ±1 yr does not necessarily mean that *all* the mass of the Homunculus was ejected within a year of 1847. An alternative possibility that would still be consistent with the kinematic data is that some of the mass loss may have occured at relatively low velocities for a prolonged period (i.e. decades) preceding the Great Eruption, but that a fast outflow that supplied most of the kinetic energy of the event occurred in the late 1840s, sweeping up all the previously ejected material into a thin shell with a single apparent age. This sort of scenario can be tested with continued study of the spectral evolution of $\eta$ Carinae's light echoes.

## ACKNOWLEDGEMENTS

I still benefit from past discussions with Jon Morse regarding *HST* imaging of Eta Carinae; many of the peculiarities we noticed in the early data are still mysterious. Support for this work was provided by NASA grants AR-12618, AR-14586, and GO-13390 from the Space Telescope Science Institute, which is operated by the Association of Universities for Research in Astronomy, Inc. under NASA contract NAS 5-26555. This work is based on observations made with the NASA/ESA Hubble Space Telescope, obtained from the Data Archive at the Space Telescope Science Institute. N.S.'s research on Eta Carinae also received support from NSF grant AST-1312221.